\documentclass[aps,prl,shownopacs,superscriptaddress,twocolumn]{revtex4}

\usepackage{graphicx}
\usepackage{hyperref}
\usepackage{amsmath}
\usepackage{epstopdf}
\newcommand{\ket}[1]{\vert #1 \rangle}
 
\begin{document}

\title{Efficient reversible entanglement transfer between light and quantum memories}

\author{Mingtao Cao \footnotemark[2]\footnotetext{\footnotemark[2]These two authors contributed equally to the work.}}
\affiliation{Laboratoire Kastler Brossel, Sorbonne Universit\'e, CNRS, ENS-Universit\'e PSL, Coll\`ege de France, 4 Place
Jussieu, 75005 Paris, France}
\affiliation{Shaanxi Province Key Laboratory of Quantum Information and Quantum Optoelectronic Devices, School of Science, Xi'an Jiaotong University, Xi'an 710049, People's Republic of China}
\author{F\'elix Hoffet \footnotemark[2]}
\affiliation{Laboratoire Kastler Brossel, Sorbonne Universit\'e, CNRS, ENS-Universit\'e PSL, Coll\`ege de France, 4 Place
Jussieu, 75005 Paris, France}
\author{Shuwei Qiu }
\affiliation{Laboratoire Kastler Brossel, Sorbonne Universit\'e, CNRS, ENS-Universit\'e PSL, Coll\`ege de France, 4 Place
Jussieu, 75005 Paris, France}
\affiliation{Shaanxi Province Key Laboratory of Quantum Information and Quantum Optoelectronic Devices, School of Science, Xi'an Jiaotong University, Xi'an 710049, People's Republic of China}
\author{Alexandra S. Sheremet\footnotemark[3]\footnotetext{\footnotemark[3]Present address: SYRTE, Observatoire de Paris, ENS-Universit\'e PSL, CNRS, Sorbonne Universit\'e, 61 Avenue de l'Observatoire, 75014 Paris, France.}}
\affiliation{Laboratoire Kastler Brossel, Sorbonne Universit\'e, CNRS, ENS-Universit\'e PSL, Coll\`ege de France, 4 Place
Jussieu, 75005 Paris, France}
\author{Julien Laurat}
\email{julien.laurat@sorbonne-universite.fr}
\affiliation{Laboratoire Kastler Brossel, Sorbonne Universit\'e, CNRS, ENS-Universit\'e PSL, Coll\`ege de France, 4 Place Jussieu, 75005 Paris, France}

%\date{\today}

\begin{abstract}
Reversible entanglement transfer between light and matter is a crucial requisite for the ongoing developments of quantum information technologies. Quantum networks and their envisioned applications, e.g., secure communications beyond direct transmission, distributed quantum computing or enhanced sensing, rely on entanglement distribution between nodes. Although entanglement transfer has been demonstrated, a current roadblock is the limited efficiency of this process that can compromise the scalability of multi-step architectures. Here we demonstrate the efficient transfer of heralded single-photon entanglement into and out-of two quantum memories based on large ensembles of cold cesium atoms. We achieve an overall storage-and-retrieval efficiency of 85\% together with a preserved suppression of the two-photon component of about 10\% of the value for a coherent state. Our work constitutes an important capability that is needed towards large scale networks and increased functionality.\\

\end{abstract}

%\setboolean{displaycopyright}{true}

\maketitle

Quantum networks rely on the transfer of quantum states of light and their mapping into stationary quantum nodes \cite{Kimble,Northup}. Central to this endeavor is the distribution of entanglement between the material nodes, which opens a variety of major applications \cite{Wehner2018,Gottesman2012,Komar2014}. For instance, for long-distance quantum communications, the distance can be decomposed into shorter quantum repeater links connecting entangled memories. Subsequent entanglement swapping operations enable an exponential improvement in distribution time \cite{Briegel98}. In this context, the efficiency of the entanglement mapping is a key parameter. As an example, an increase in storage-and-retrieval efficiency from 60\% to 90\% drastically decreases -- typically by two orders of magnitude -- the average time for entanglement distribution over a distance of 600 kilometers \cite{Sangouard2011}. 

\begin{figure}[b!]
\includegraphics[width=0.96\columnwidth]{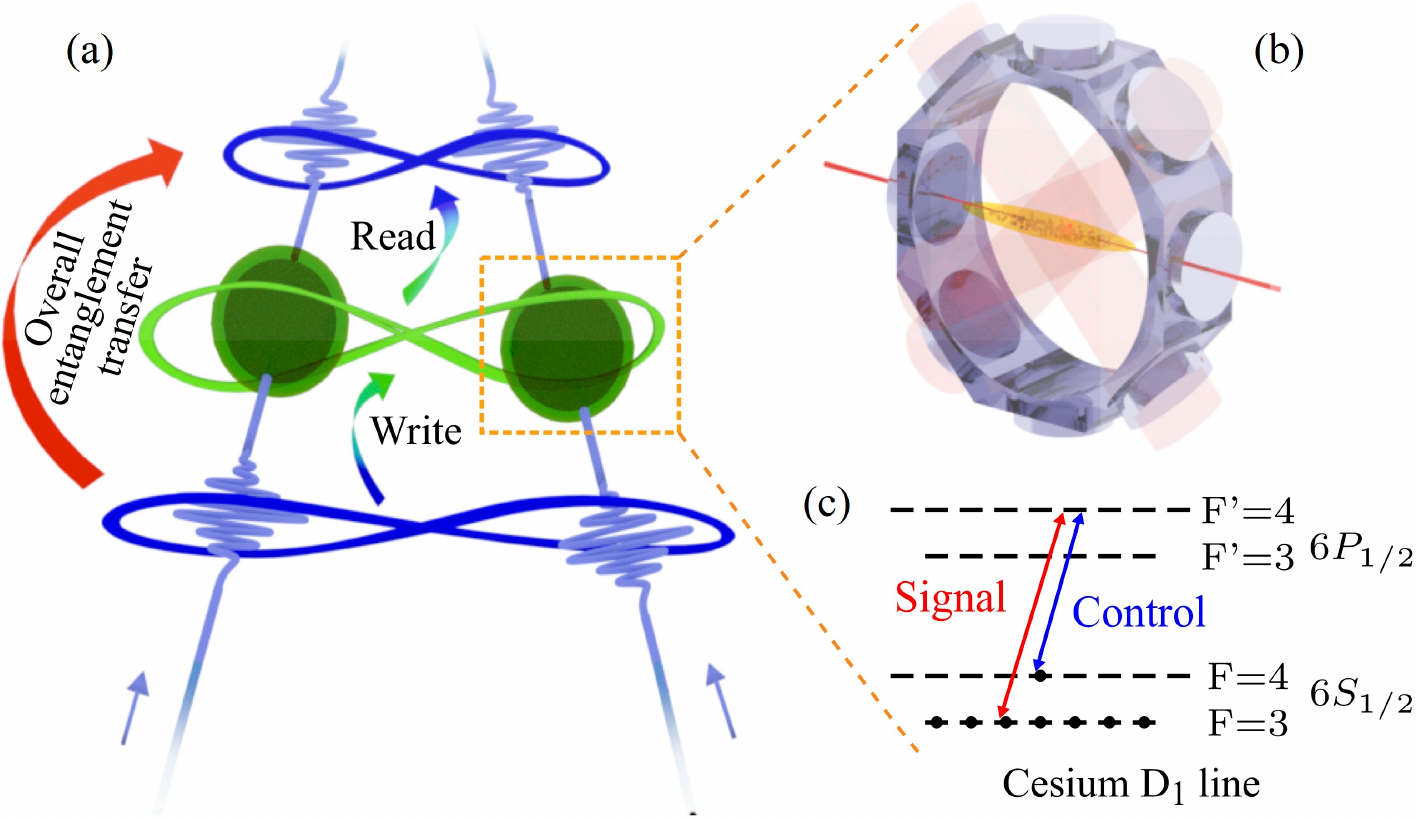}
\caption{Reversible entanglement transfer. (a) Single-photon entanglement is heralded, stored into two quantum memories and read out on demand. The overall efficiency of the writing and reading transfer is a key parameter for scaling up quantum networks. (b) The memories are based on elongated ensembles of cold cesium atoms. (c) The EIT scheme used for storage is implemented on the cesium D$_1$ line. The entangled fields are resonant with the $\ket{g}~=~\ket{6S_{1/2}, F=3}$ to $\ket{e} = \ket{6P_{1/2}, F'=4}$ transition while the control field is tuned with the $\ket{s} = \ket{6S_{1/2}, F=4}$ to $\ket{e}$ transition.}
\label{fig1}
\end{figure}

In this endeavor, quantum state transfer and entanglement mapping between photonic modes and stationary quantum nodes has been demonstrated in different physical platforms \cite{Heshami2016, Hammerer2010,Rempe,Dayan}. Seminal experiments based on quantum memories with cold neutral atoms \cite{Choi2008,Ding2015} or doped crystals \cite{Usmani2012} have enabled the storage and retrieval of heralded single-photon entanglement. Extensions towards high-dimensional \cite{Zhou2015} and continuous-variable entanglement \cite{Jensen2011,Yan2017} have also been reported. However, in all these implementations the overall transfer efficiency was limited between 15\% and 25\%. Despite the recent demonstrations of efficient quantum memories for polarization qubits \cite{VernazGris2018,Wang2019}, efficient entanglement transfer is a major challenge for network scalability that has yet to be realized.

Here, we demonstrate the implementation of highly-efficient and reversible entanglement transfer combined with a very low multiphoton component. As illustrated in Fig. \ref{fig1}, single-photon entanglement is first heralded and then stored into two quantum memories based on elongated atomic ensembles of cold cesium atoms. After readout, entanglement is detected and compared to the input. Our implementation relies on temporally-shaped single-photon pulses generated via the Duan-Lukin-Cirac-Zoller protocol (DLCZ) \cite{DLCZ2001,Laurat07a} and on the dynamic electromagnetically induced transparency (EIT) technique for reversible storage \cite{Lukin}. The demonstrated capability required operating at a very large optical depth (OD) of the atomic ensembles on the D$_1$ line of cesium, and with a strong and preserved suppression of the two-photon component.

\section{Experimental implementation}
\begin{figure}[t!]
\centering
\includegraphics[width=0.99\columnwidth]{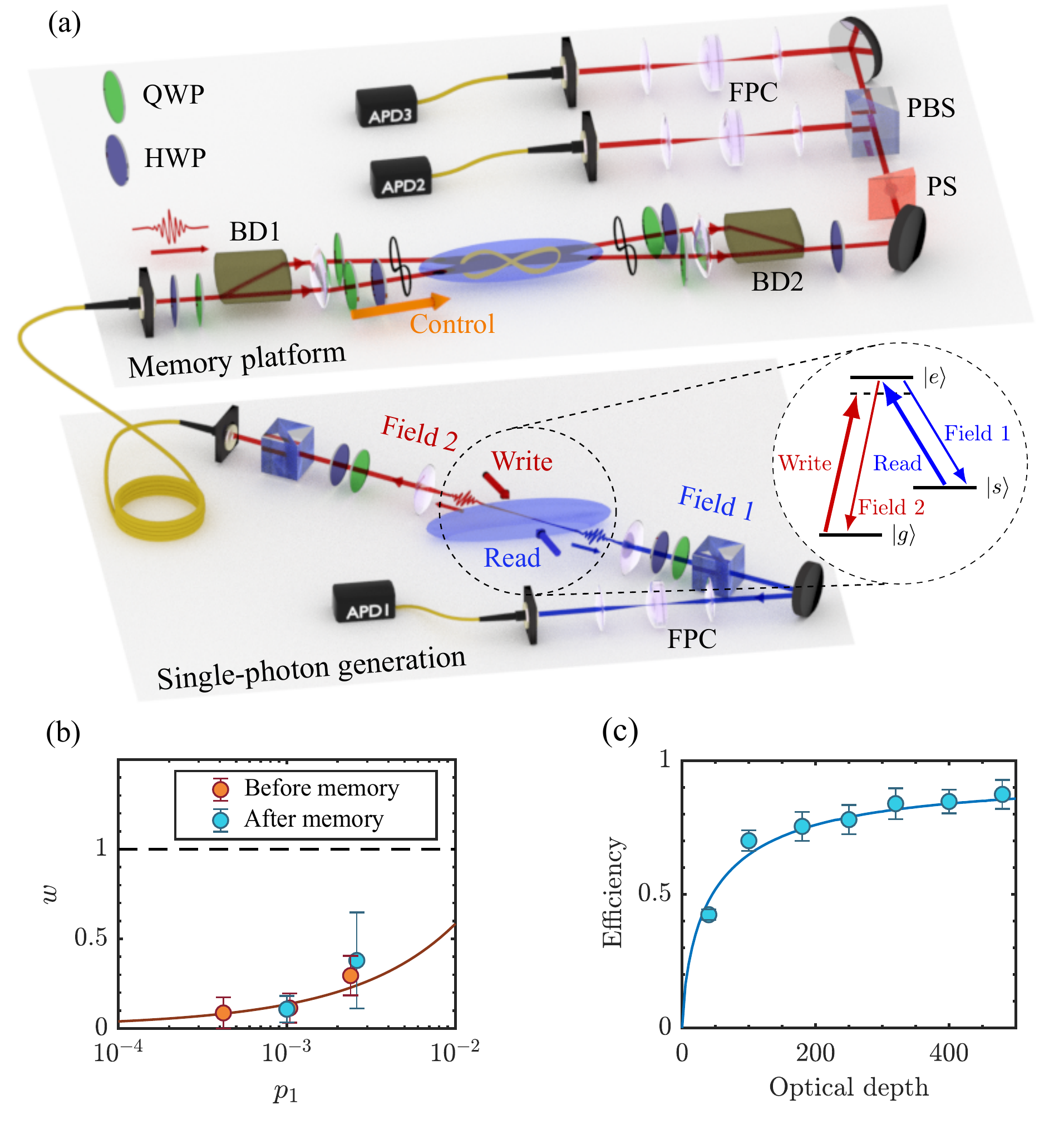}
\vspace{-0.5cm}
\caption{Setup and memory characterization. The experiment consists on two stages operated with the same ensemble of Cs atoms. (a) A 2.5-cm-long ensemble is prepared in a compressed MOT, with an OD up to 500. At first, a single photon is generated using the DLCZ protocol in a small transverse part of the ensemble. After sending a write pulse, the detection of a Field-1 photon, with a probability $p_1$, heralds the creation of a single collective excitation. A read pulse produces then a Field-2 single photon that propagates in a 1-$\mu$s fibered delay line and impinges on a beam displacer (BD). The resulting single-photon entanglement is mapped into and out-of two memory ensembles via EIT, with a control propagating at $1^{\circ}$ angle. Entanglement is finally characterized using a phase shifter (PS), a polarizing beamsplitter (PBS) and two APDs. Before detection, Fabry-Perot cavities (FPC) filter the control background. (b) Suppression $w$ of the two-photon component for Field 2 relative to a coherent state before and after retrieval as a function of $p_{1}$. (c) Storage-and-retrieval efficiency for a single photon in one of the memories as a function of OD. The efficiency reaches $(87\pm5)\%$. The line corresponds to the full model (see text). The error bars are obtained from the Poissonian error of the photon counting probabilities.}
\label{fig2}
\end{figure}

The setup is detailed in Fig. \ref{fig2}(a) and relies on a single 2.5 cm-long atomic ensemble. At a 20 Hz repetition rate, the experimental runs start with a loading phase of 37.5~ms in the magneto-optical trap. To achieve a large OD, the trap is based on two pairs of rectangular coils and on 2-inch-diameter trapping beams with a total power of 350~mW. An additional 8-ms compression stage, with ramping of the trapping coils current from 3.5 to 15~A, is performed. The MOT coils are then switched off and polarization gradient cooling is performed for 1~ms. Finally, the atoms are prepared in the ground state $\ket{g}$ by sending a 950-$\mu$s long pulse resonant with the $\ket{s}$ to $\ket{e}$ transition (see Fig. \ref{fig1}). Residual magnetic fields are cancelled and the inhomogeneous broadening of the hyperfine ground state transition is limited to 50 kHz. Overall, the OD on the signal transition reaches 500. 

After the preparation of the atomic sample, the experiment consists on two stages, performed subsequently 250 times per MOT cycle.  A single photon is first generated based on the DLCZ protocol \cite{DLCZ2001,Laurat07a} operated on a small transverse part of the elongated ensemble, with an OD of 6. A 150-ns write pulse detuned by -15~MHz with the $\ket{g} $ to $\ket{e}$ transition induces spontaneous Raman-scattered fields. The detection of a Field-1 photon with the single-photon detector APD1, with a probability $p_1$, heralds the creation of a collective atomic excitation. After a programmable delay, set at 200~ns, a pulse resonant with the $\ket{s}$ to $\ket{e}$ transition reads out this excitation and a field-2 single photon is emitted into a well-defined spatio-temporal mode. To match the spectral-temporal properties of the subsequent EIT storage, the single photon is temporally shaped via the shaping of the read pulse \cite{Farrera2016}. With a 500-ns Gaussian read pulse, a 300-ns Gaussian-shaped single-photon pulse is obtained. Due to the limited OD used for this stage, the heralding efficiency, i.e., the probability to get a single photon out of the ensemble, is equal to 10\%.We note that a larger OD would enable to reach a larger efficiency, with values up to 50\% demonstrated in free-space experiments with OD about 20 \cite{Laurat07a}. However, with an angle between the fields as generally used for filtering and an elongated atomic ensemble necessary for a large OD, the phase-matching mismatch due to the splitting in the ground state can effectively limit the achievable efficiency. This limitation is not present for the EIT process due to the different configuration of the beams. 

To characterize the emitted single photon, the suppression $w$ of the two-photon component relative to a coherent state is measured via a Hanbury Brown-Twiss setup. It is given by the ratio $p_{1}p_{1,2,3}/(p_{1,2} p_{1,3})$, where $p_{1,2,3}$  indicates the probability for triple coincidences and $p_{1,2}$ ($p_{1,3}$) the probability for coincidences between APD1 and APD2 (APD3). This antibunching parameter depends on the excitation probability $p_{1}$ and the values are given in Fig. 2(b).

The single photon is then sent through a 200-meter long single-mode fiber that introduces a 1-$\mu$s delay. After this propagation, the single photon impinges on a birefringent beam displacer (BD1) with a polarization at $45^{\circ}$ from the axis, generating in the ideal case the single-photon entangled state 
\begin{eqnarray}
\frac{1}{\sqrt{2}}\left ( \ket{0_{a}}\ket{1_{b}}+e^{i\varphi}\ket{1_{a}}\ket{0_{b}} \right)\nonumber
\label{entangle in}
\end{eqnarray}
where $a$ and $b$ denote the two optical paths. The relative phase $\varphi$ is passively stable due to the small interferometer defined by the BDs \cite{Matsukevich2004,Chou07,Laurat07}. Entangling remote memories would require active phase locking, as demonstrated in \cite{Chou2005}. The focus of our study is here on the efficiency of the entanglement transfer that can be achieved.

The second stage of our experiment consists in coherently mapping the entangled fields into two memories by dynamical EIT. The ensembles are defined by two crossed optical paths, which are obtained by focusing the two 4-mm-apart parallel paths into the MOT with a small angle of 0.5$^{\circ}$ and a waist of 250 $\mu$m. A control field, on resonance with the $\ket{s}$ to $\ket{e}$ transition and with a waist of 3 mm, opens a transparency window in both atomic ensembles. When the entangled fields propagate through the ensembles, the control is switched off adiabatically to coherently map them onto long-lived collective excitation, leading here to one excitation delocalized among the two memories. After a tunable delay, set to 1~$\mu$s for the presented results, the atomic entanglement is converted back into entangled photonic modes by switching on the control. Both control and entangled fields have the same circular polarization. To avoid leakage during the mapping, i.e., to ensure that the signal can be contained entirely in the ensemble when the control is switched off, the control power is chosen to provide a slow-light delay equal to twice the signal duration. In order to achieve an efficient entanglement transfer, a large storage-and-retrieval efficiency and therefore a very high OD is required for both path. The small cross angle enables to preserve a value up to 500. However, as studied in \cite{VernazGris2018,Hsiao2018}, the off-resonance excitation of multiple excited levels in alkaline atoms result into an effective decoherence rate that can limit the efficiency for large OD. For this reason, the experiment is performed here on the D$_1$ line, where the separation between the two excited states is larger than 1.1~GHz and this effect is thereby minimized relative to experiments on the D$_{2}$ line for which efficiency peaks at about 70\% before decreasing \cite{VernazGris2018}.

\section{Memory characterization}
We first study the mapping of the heralded single photon into and out of one memory, using one path of the interferometer. Figure \ref{fig2}(c) displays the measured storage-and-retrieval efficiency as a function of OD. Similar results are obtained for both memories. The efficiencies are compared to a full model based on Maxwell-Bloch equations, which takes into account the interaction of the signal and control with all the Zeeman and excited levels \cite{VernazGris2018}. For this model, we consider an intrinsic ground-state decoherence rate $\gamma_0=10^{-3} \Gamma$, as extracted from EIT spectra measurement, where $\Gamma/2\pi=4.5$~MHz is the decay rate for D$_1$ line. The data agree well with this model. The maximal achieved efficiency reaches (87$\pm$5)\% for an OD of about 500. As can be seen in Fig. \ref{fig2}(b) that provides the suppression $w$ of the two-photon component before and after storage, the single-photon character is very well preserved. For the lowest excitation probability $p_1\simeq10^{-3}$ used in the memory experiment, $w$ is equal to $0.11\pm0.08$ and shows no degradation within the error bar. Relative to the only other work on single-photon storage with high efficiency \cite{Wang2019}, this value is here three times lower, which is a stringent requirement for quantum repeater applications \cite{Chou07,njp}. 

In this implementation, the memory lifetime $\tau$, given by a gaussian decay $\exp\left(-(t/\tau)^2\right)$ as it comes from the residual inhomogeneous Zeeman broadening \cite{Choi}, is measured to be $15~\mu s$. Motional dephasing due to the angle between the signal and the control would otherwise limit the lifetime to about 200~$\mu s$ while the transit of the atoms from the interaction area when the MOT is released puts an upper limit below 10~ms. Various improvements could lead to a few-millisecond timescale, e.g. reducing the angle, which will require a more efficient filtering, optical pumping, which is challenging in high-OD media, or magnetic field bias to lift Zeeman degeneracy \cite{Xu2013}. To access the sub-second regime, other optical trapping methods, e.g, dipole trapping, are required as demonstrated in Refs. \cite{Radnaev2008,Yang2016} with 3D optical lattices. In that case, a cavity around the atomic ensemble might be necessary as in Ref. \cite{Yang2016} to preserve a large OD. 

For completeness, we give here the typical experimental rates. During the 1-ms phase when generation and storage are performed, the heralding rate to generate the single photon is of about 25 per second and the rate of entanglement generation is reduced to 18 per second due to couplings and loss in the delay line. After storage-and-retrieval, propagation and filtering (30\% transmission), and detection (50\% efficiency), the single-photon detection rate is of about 1.7 per second. Given the specific duty cycle of 1/50 in our implementation, the overall entanglement generation rate and single-photon detection rate are of about 0.3 and 0.03 per second, respectively.

\section{Entanglement transfer}

We now turn to the entanglement characterization. For this purpose, we follow the model-independent determination introduced in \cite{Chou2005} that consists of measuring a reduced density matrix $\tilde{\rho}$ by restricting to the subspace with no more than one photon per mode and assuming that all off-diagonal elements between states with different numbers of photons are zero. This method provides a lower bound for the entanglement. In the basis $\ket{i_{a},j_{b}}$ with the number of photons $\{i,j\}=\{0,1\}$, $\tilde{\rho}$ can be written as
\begin{eqnarray}
\tilde{\rho} = \frac{1}{P}
\begin{pmatrix}
p_{00}&0&0&0\\
0&p_{01}&d&0\\
0&d^*&p_{10}&0\\
0&0&0&p_{11}
\label{matrix}
\end{pmatrix}\nonumber
\end{eqnarray}
where $p_{i,j}$ corresponds to the probability to find $i$ photon in mode $a$ and $j$ photon in mode $b$, $P=p_{00}+p_{01}+p_{10}+p_{11}$, and d is the coherence between the states $\ket{0_{a},1_{b}}$ and $\ket{1_{a},0_{b}}$. The coherence term is given by $d=V(p_{01}+p_{10})/2$ where $V$ is the visibility of the interference fringe between mode $a$ and $b$ when their relative phase is scanned. The reduced density matrix enables to calculate the concurrence $\mathcal{C}$ \cite{Wooters}, i.e., a monotone measurement of entanglement, as
\begin{eqnarray}
    \mathcal{C}=\frac{1}{P}\textrm{max}\left(2d-2\sqrt{p_{00}p_{11}},0\right)\nonumber
\end{eqnarray}
where $\mathcal{C}$ takes values between 0 for a separable state to 1 for a maximally entangled state.

\begin{figure}[t]
\centering
\includegraphics[width=0.96\columnwidth]{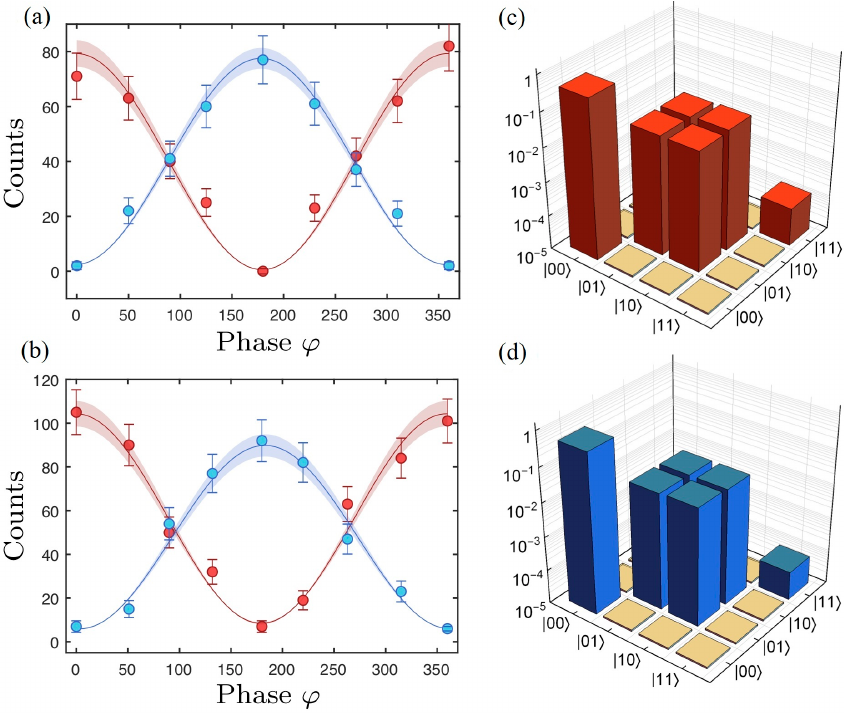}
\caption{Measurement of entanglement for the input and output entangled fields. The relative phase $\varphi$ between the two modes is scanned and lead to interferences fringes for (a) before storage with visibility $V=0.96\pm0.03$ and (b) after storage and read-out with $V=0.87\pm0.04$ ($V=0.94\pm0.04$ after background correction). The density matrix is then derived for (c) before the memory with a concurrence $\mathcal{C}_{\textrm{in}}=(5.9\pm1.2)\times10^{-3}$ and (d) after the memory with $\mathcal{C}_{\textrm{out}}=(4.7\pm0.9)\times10^{-3}$. The error bars correspond to the propagated Poissonian error of the photon counting probabilities and the $1\sigma$ confidence interval of the sine fit applied to the fringes.}
\label{fig3}
\end{figure}

Experimentally, upon the detection of a heralding photon on APD1, we first measure the $p_{i,j}$ probabilities with APD2 and APD3. The two detectors assess the presence of photons in mode $a$ or $b$ by monitoring the two outputs of a polarizing beam splitter placed after recombination of the two paths of the interferometer. The measured probabilities are provided in Table \ref{table1}. The next step is to measure interference fringes by mixing the two modes and scanning their relative phase $\varphi$. This can be done by using a set of two half-wave plates placed after BD2, with their axis parallel to the field polarizations and varying their relative angle. The experimental fringes are given in Fig. \ref{fig3}(a) and \ref{fig3}(b). The average raw visibilities are $V_{\textrm{in}}=0.96\pm0.03$ and $V_{\textrm{out}}=0.87\pm0.04$. The decrease in visibility after storage is mainly due to a slight contamination by the control field. After correction of this background we obtain a visibility $V_{\textrm{out}}=0.94\pm0.03$. The reconstructed density matrices $\tilde{\rho}_{\textrm{in}}$ and $\tilde{\rho}_{\textrm{out}}$  are given in Fig. \ref{fig3}(c) and \ref{fig3}(d), respectively.

We note that when measuring the entanglement for the input modes, we could not remove the atoms since we use the same atomic ensemble to generate the single photon and store subsequently the entanglement. Therefore, we built another path with an identical interferometer next to our atomic cloud. This reference path is recombined with the memory path before filtering and detection. We estimate the input entanglement by correcting on the losses ratio between the two interferometers.

\setlength{\tabcolsep}{3.7mm}{
\begin{table}[b!] 
 \caption{Measured probabilities ${p_{ij}}$ and estimated concurrences $\mathcal{C}$ before and after storage, without correction for loss, detection and noise. The error bars correspond to the propagated Poissonian error of the photon counting probabilities.} 
 \begin{tabular}{lll} 
  \hline\hline
   & ${{\tilde{\rho} }_{\textrm{in}}}$ & ${{\tilde{ \rho} }_{\textrm{out}}}$ \\ 
  \hline
 ${{ p}_{00}}$ & $0.991 \pm 0.001$& $0.992 \pm 0.001$ \\ 
 ${{ p}_{10}}$ & $\left( {4.57 \pm 0.12} \right) \times {10^{ - 3}}$ & $\left( {3.87 \pm 0.09} \right) \times {10^{ - 3}}$ \\ 
 ${{ p}_{01}}$ & $\left( {4.95 \pm 0.12} \right) \times {10^{ - 3}}$ & $\left( {4.18 \pm 0.09} \right) \times {10^{ - 3}}$ \\ 
 ${{ p}_{11}}$ & $\left( {2.58 \pm 1.80} \right) \times {10^{ - 6}}$ & $\left( {1.35 \pm 0.95} \right) \times {10^{ - 6}}$ \\ 
 $\mathcal{C}$ & $\left( {5.9 \pm 1.2} \right) \times {10^{ - 3}}$ & $\left( {4.7 \pm 0.9} \right) \times {10^{ - 3}}$ \\ 
 \hline\hline
 \end{tabular} 
\label{table1}
\end{table}}

The raw data given in Table \ref{table1} allow to characterize the entanglement transfer performances. The first crucial parameter is the suppression of the two-photon component. It can be evaluated here by the ratio $w=p_{11}/(p_{10}\cdot p_{01})$. This parameter amounts to $w_{\textrm{in}}=0.11\pm0.07$ and $w_{\textrm{out}}=0.08\pm0.06$ before and after storage respectively, thereby confirming the preservation of the single-photon character. Also, the overall storage-and-retrieval efficiency $\eta$ is given by the ratio of the one-photon probabilities $(p_{10}+p_{01})_{\textrm{out}}/(p_{10}+p_{01})_{\textrm{in}}$. This ratio is equal to $\eta=(85\pm4)\%$, in agreement with the efficiencies measured for each memory operated independently. These values combined with the achieved visibilities confirm the efficient, noiseless and reversible coherent mapping.   

From these data, one can estimate the concurrence of entanglement. Without correcting for losses, detection efficiencies and residual background noise, we obtain a value for the concurrence before storage of $\mathcal{C}_{\textrm{in}}=(5.9\pm1.2)\times10^{-3}$ and a value after retrieval of  $\mathcal{C}_{\textrm{out}}=(4.7\pm0.9)\times10^{-3}$. The entanglement transfer can be evaluated by the ratio of the concurrences, $\lambda~=~\mathcal{C}_{\textrm{out}}/\mathcal{C}_{\textrm{in}}$, as initially done in \cite{Choi2008}. In our implementation, this parameter reaches $\lambda=(80\pm20)\%$. In the ideal case, this value is equal to the storage-and-retrieval efficiency of the involved memories. If the visibility of the fringes obtained after retrieval is corrected from background noise, the output concurrence is increased to $(5.3\pm0.9)\times10^{-3}$ and the ratio $\lambda$ to $88^{\tiny{+12}}_{\tiny{-23}}\%$. These numbers represent more than a 3-fold increase in transfer efficiency relative to previous works. 

The error bar on $\lambda$ is mainly due to the uncertainty obtained on the values of $p_{11}$. This parameter is known to be difficult to measure as it corresponds to events for which the occurrence decreases rapidly with the suppression of the two-photon component \cite{Choi2008}. 180 hours of data taking were necessary to specifically access these probabilities and few coincidences were obtained. Indirect methods could be used to assess entanglement but would require specific assumptions about the initial state and noise statistics \cite{Usmani2012} or performing homodyne measurements \cite{Sangouard}. In the broad context of quantum networks, this result also emphasizes the topical importance of developing efficient benchmarking tools \cite{Eisert}. 

\section{Conclusion}
In conclusion, we have reported the first realization of a highly-efficient and reversible entanglement transfer between light and quantum memories, together with a strong and preserved suppression of the two-photon component. The demonstrated capability is an important step towards the development of scalable networking architectures. A central challenge remains to demonstrate a \textit{quantum link efficiency} greater than unity \cite{Hanson,Laurat}, i.e., a preparation rate of entangled memories much larger than the decoherence rate, a cornerstone yet to be demonstrated in a cold atom setting. Combining high efficiency as shown here with longer lifetime \cite{Xu2013,Radnaev2008,Yang2016} and with multiplexing in multiple degrees of freedom \cite{Nicolas2014,Pu2017,Parniak2017,Simon2010} will be necessary. \\

This work was supported by the European Research Council (Starting Grant HybridNet), the EU H2020 FETFLAG Quantum Internet Alliance (820445), and the French National Research Agency (quBIC project ANR-17-CE39-0005). M.C. and A.S.S. acknowledge the support from the EU (Marie Curie Fellowship projects 705161 and 708216) and S.Q. from the China Scholarship Council. We also thank P. Vernaz-Gris and K. Huang for their contributions in the early stage of the experiment.

%% Full list of references

\end{document}